\documentclass[aip,jcp,superscriptaddress,showpacs,preprint]{revtex4}
\usepackage{amsfonts}
\usepackage{amsmath}
\usepackage{amssymb}
\usepackage{graphicx}
\usepackage[T1]{fontenc}
\usepackage[latin9]{inputenc}
\usepackage{array}
\usepackage{longtable}
\usepackage{amsmath}
\usepackage{esint}

\providecommand{\U}[1]{\protect\rule{.1in}{.1in}}

\begin{document}
\title{Quantum molecular dynamics simulations for the nonmetal-metal transition
in shocked methane}
\author{Dafang Li}
\affiliation{LCP, Institute of Applied Physics and Computational
Mathematics, Beijing 100088, People's Republic of China}
\author{Ping Zhang}
\thanks{zhang\_ping@iapcm.ac.cn}
\affiliation{LCP, Institute of Applied Physics and Computational
Mathematics, Beijing 100088, People's Republic of China}
\affiliation{Center for Applied Physics and Technology, Peking
University, Beijing 100871, People's Republic of China}
\author{Jun Yan}
\thanks{yan\_jun@iapcm.ac.cn}
\affiliation{LCP, Institute of Applied Physics and Computational
Mathematics, Beijing 100088, People's Republic of China}
\affiliation{Center for Applied Physics and Technology, Peking
University, Beijing 100871, People's Republic of China}

\pacs{65.20.De, 64.30.Jk, 51.70.+f, 31.15.xv}

\begin{abstract}
We have performed quantum molecular-dynamics simulations for methane
under shock compressions up to 80 GPa. We obtain good agreement with
available experimental data for the principal Hugoniot, derived from
the equation of state. A systematic study of the optical
conductivity spectra, one-particle density of states, and the
distributions of the electronic charge over supercell at Hugoniot
points shows that the transition of shocked methane to a metallic
state takes place close to the density at which methane dissociates
significantly into molecular hydrogen and some long alkane chains.
We predict the chemical picture of the shocked methane with respect
to the pair correlation function . In contrast to usual assumptions
used for high pressure modeling of methane, we find that no
diamond-like configurations occurs for the whole density-temperature
range studied.
\end{abstract}
\maketitle

\section{INTRODUCTION}

The nature of methane under extreme conditions has recently drawn
extensive attentions due to the ongoing scientific and technological
interest. As methane is one major constituent of the {}``ice''
layers in Uranus and Neptune, where the pressure ranges between 20
and 600 GPa and temperature between 2000 to 8000 K
\cite{Hubbard1980,Stevenson1982}, its properties are of considerable
impact on the internal evolution and energetics of these giant
planets. The knowledge of its equation of state and electrical
conductivity at these conditions are necessary both for theoretical
modeling of the composition of these planets and for understanding
its contribution to the planetary magnetic fields, which are caused
by convective dynamo action of electrically conducting fluid
\cite{Stevenson1983}. Thus, it is desirable to determine the
transformations in methane induced by nonequilibrium phenomena such
as shock waves and detonations.

A lot of efforts have been devoted to exploring the methane chemical
dissociation under extreme conditions. At high static pressures
between 10 and 50 GPa and temperatures of about 1000 to 3000 K,
diamond-anvil cell experiment has suggested that methane breaks down
to form diamond and hydrogen \cite{LBenedetti1999}, in qualitative
agreement with Ree's prediction \cite{Ree1979}. However, on the
methane Hugoniot \cite{Nellis1981} different results were reported.
No dissociation was found in shocked methane for compressions up to
42 GPa \cite{Radousky1990}, while Nellis et al. observed the
increase of electrical conductivity in methane shocked by two-stage
light-gas gun, which was attributed to the decomposition into a
substantial concentration of molecular hydrogen at 36 GPa and 3900 K
\cite{Nellis2001}. The discrepancy may be caused by the shorter time
scale of shock wave experiments. Concerning the theoretical side,
empirical potentials \cite{Elert2003,Bolesta2007} and tight-binding
simulations \cite{Kress1999} have been used to conduct molecular
dynamics simulations of the shock compression of methane, while the
use of density functional theory in first-principles molecular
dynamics simulations \cite{Ancilotto1997,Goldman2009} provides more
accurate modeling of the breaking and forming of chemical bonds.
FPMD simulations with 16 methane molecules for 2 picoseconds show
that some polymerization reactions occur at 100 GPa and 4000 K and
that diamond formation take place only above 300 GPa
\cite{Ancilotto1997}. Recent ab initio evolutionary simulations
\cite{Gao2010} have systematically investigated the phase diagram of
methane under pressure, and confirmed the dissociation of methane at
high pressures. Whereas, the anharmonism was neglected in
calculations, which may affects the stability of methane.

To date, although a number of explanatory and predictive results in
some cases have already been provided by experimental and
theoretical studies, many fundamental questions of methane at
extreme conditions are still much controversial and under debate.
The systematic study of equation of state, especially the Hugoniot
curve under extreme conditions is far lacking now, which is
essential in this context. Furthermore, in adiabatic compressions,
the nonmetal-metal transition of hydrocarbons has been a major issue
recently. For example, benzene \cite{Wang2010} has been reported to
transform metallic phase when decomposing into hydrogen under
extreme conditions. To date there are still no such data explored
for methane, especially due to the difficulties in applying current
opacity models at these particular conditions. It is thus highly
needed to study the electronic structure, Hugoniot EOS, and optical
properties of shocked methane using an efficient method
consistently. Quantum molecular dynamics (QMD), where the active
electrons are treated in a full quantum mechanical way within the
finite temperature density functional theory (FT-DFT), has been
proven a successful tool to calculate the properties of complex
plasmas under such extreme conditions
\cite{Collins2001,Mazevet2003,Kress2001,Desjarlais2002,Mazevet2004,Laudernet2004}.
When combined with the Kubo-Greenwood formulation, the method
produces a consistent set of material, electrical, and optical
properties from the same simulation and can be applied to various
systems.

In the present work, we perform QMD simulations for the system with
densities and temperatures ranging from 0.6
$\textnormal{g/c\ensuremath{m^{3}}}$ and 175 K to 1.5
$\textnormal{g/c\ensuremath{m^{3}}}$ and 7000 K along the principal
Hugoniot on an intermediate time scale. The analysis of the
concentration of molecular species along the Hugoniot based on pair
correlation function (PCF) indicates that methane decomposes into
molecular hydrogen and saturated hydrocarbons first, and then into
long alkane chains at higher pressures and temperatures, which
consequently leads to the nonmetal-metal transition. The dynamic
conductivity, calculated using the Kubo-Greenwood formula, along
with electron density of states and charge distribution, confirms
the occurrence of nonmetal-metal transition around 55 GPa. This
paper is organized as follows. The simulation details are briefly
described in Sec. II; the PCF, which is used to study the
dissociation of methane, and Hugoniot curve are given in Sec. III;
in Sec IV nonmetal-metal transition, optical properties and
electronic properties are discussed. Finally, we close our paper
with a summary of our main results.

\section{COMPUTATIONAL METHOD}

In the present application, the molecular dynamics trajectories are
calculated by employing the Vienna ab initio simulation package
(VASP) plane-wave pseudopotential code developed at the Technical
University of Vienna \cite{Kresse1993,Kresse1996}. In these
simulations, the electrons receive a fully quantum mechanical
treatment through solving the Kohn-Sham (KS) equations for a set of
orbitals and energies within a plane-wave, FT-DFT formulation
\cite{Lenosky2000,Bagnier2001}, where the electronic states are
populated according to the Fermi-Dirac distribution at temperature
$T_{e}$. The all-electron projector augmented wave (PAW) method
\cite{Kresse1999,Bloch1994} is adopted and the exchange-correlation
energy is described employing the Perdew-Wang 91 parametrization of
the generalized gradient approximation (GGA) \cite{Perdew1991}.
Atoms move classically according to the forces, which originate from
the interactions of ions and electrons.

We perform finite-temperature, fixed-volume molecular dynamics
simulations for selected densities ranging from 0.6 to 1.5
$\textnormal{g/c\ensuremath{m^{3}}}$ and temperatures from 175 to
7000 K that highlight the single-shock Hugoniot region. We use 27
carbon and 108 hydrogen atoms (twenty-seven methane molecules) in a
cubic cell and fix the plane-wave cutoff at 600.0 eV which is tested
to give good convergence for both total energy and pressure. The
Brillouin zone is sampled with the $\Gamma$ point for molecular
dynamics and $4\times4\times4$ Monkhorst-Pack \cite{Monkhorst1976}
scheme $k$ points for the electronic structure calculations.
Integration of the equations of motion proceeds with time steps of
0.5-1.0 fs for different pressure-temperature ranges. Typical
simulations run for 8000-16000 steps with the time scale about 8 ps.
We let the system equilibrate for 4000-8000 steps and calculate
properties using the final 4000-8000 steps. The isokinetic ensemble
(NVT) is employed for the ions, where the ion temperature $T_{i}$ is
fixed using velocity scaling. The electron temperature $T_{e}$ is in
turn set to that of the ions $T_{i}$ based on the assumption of
local thermodynamical equilibrium. The accuracy of our calculations
is examined by the bond length of CH4 molecule in its ground state
and the result is 1.09, which agrees well with the experiment.

\section{SHOCK EOS AND PCF}

A precise description of material properties, such as EOS, is
important to the accurate calculation of electrical and optical
properties. A crucial measure for theoretical EOS data is the
principal Hugoniot curve. It describes the shock adiabat through a
relation between the initial and final internal energy, pressure and
volume, respectively, $\left(E_{0},\: P_{0},\: V_{0}\right)$ and
$\left(E_{1},\: P_{1},\: V_{1}\right)$ according to the following
Rankine-Hugoniot equation \cite{Zeldovich1966}:

\begin{eqnarray}
\left(E_{1}-E_{0}\right)+\frac{1}{2}\left(V_{1}-V_{0}\right)\left(P_{0}+P_{1}\right)
& = & 0,\end{eqnarray} where the internal energy $E$ equals to the
sum of the ion kinetic energy $\frac{3}{2}k_{B}T_{i}$, the time
average of the DFT potential energy and zero-point energy with
$k_{B}$ the Boltzmann constant. The pressure consists of
contributions from the electronic $P_{e}$ and ionic $P_{i}$
components, which come from, respectively, the derivatives taken
with respect to the KS electronic orbitals and the ideal gas
expression since ions move classically. We thus have
$P=P_{e}+\rho_{n}k_{B}T$, where $\rho_{n}$ is the number density.
For the methane principal Hugoniot curve, the initial density is
$\rho_{0}=0.423\:\textnormal{g/c\ensuremath{m^{3}}}$ and the initial
internal energy $E_{0}=23.30$ eV/molecule at a temperature of 111 K.
The initial pressure can be neglected compared to the high pressure
of the final state. The Hugoniot points are determined in the
following way. For a given density, the periodic crystalline sample
of the corresponding size was first constructed. A fixed volume
cubic supercell of 27 carbon and 108 hydrogen atoms (27 methane
molecules), which is repeated periodically throughout the space,
forms the elements of the calculation. A series of quantum molecular
dynamics simulations are performed for different temperatures $T$.
Following it, the internal energies and pressures are determined
correspondingly and then fitted to a cubic function of $T$. For a
given density and a set of temperatures, we plotted
$(V_{0}-V_{1})(P_{0}+P_{1})/2$ along with $(E_{1}-E_{0})$ as a
function of temperature. The intersection of the two curves fixes
the principal Hugoniot point $\left(E_{1},\: P_{1},\: V_{1}\right)$
that satisfies Eq.(1). The particle velocity $u_{p}$ and shock
velocity $u_{s}$ are then determined from the other two
Rankine-Hugoniot equations \cite{Zeldovich1966},
$V_{1}=V_{0}\left[1-\left(u_{p}/u_{s}\right)\right]$ and
$P_{1}-P_{0}=\rho_{0}u_{s}u_{p}$. The principal Hugoniot points of
methane derived from Eq. (1) are listed in Table I.

\begin{figure}
\includegraphics[scale=0.50]{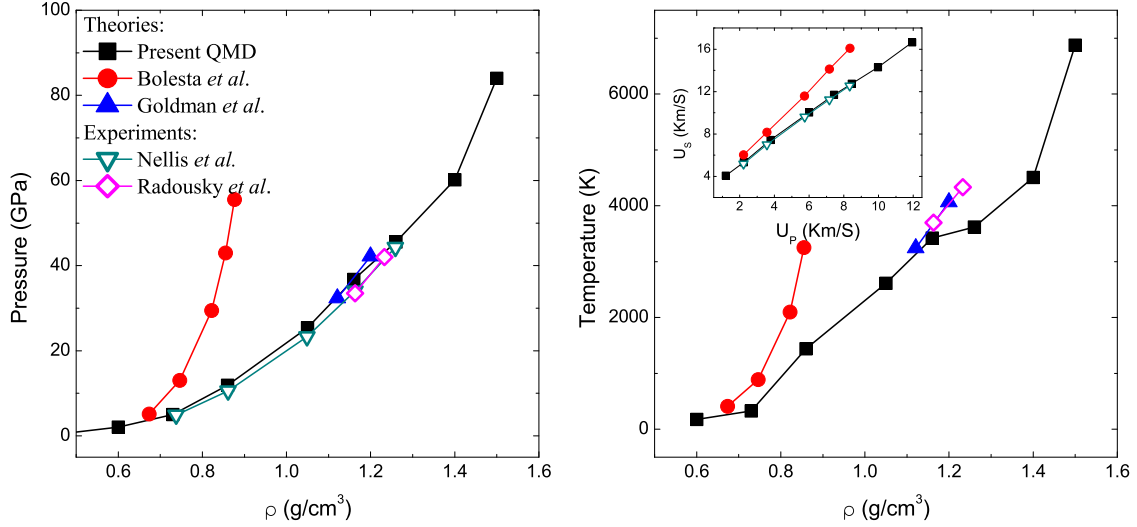}

\caption{Principal Hugoniot of methane. For comparison, the present
QMD results, together with the previous experimental data (Ref. 6
and Ref. 7) and other theoretical results (Ref. 10 and Ref. 13) are
all provided.}

\end{figure}

We display our simulated Hugoniot curve for methane in Fig. 1, along
with experimental data and results from previous theoretical works.
From Fig. 1 we find very good agreement with the experimental
results all along the single-shock Hugoniot, while the data from the
empirical potential simulations based on the adaptive intermolecular
reactive empirical bond order force field presents prominent
discrepancy with all the other results, which may be caused by the
inaccurate modeling of the breaking and forming of chemical bonds.
Our simulated pressures and temperatures using QMD show a systematic
behavior along the Hugoniot curve, except for the region with a
small break in slope of temperature plot between 1.16 and 1.26
$\textnormal{g/c\ensuremath{m^{3}}}$, which usually implies the
dissociation of molecular species in the media and thus suggests
that methane is dissociating in this density region.

\begin{table}
\caption{Points along the principal methane Hugoniot derived from
DFT-MD simulations at a series of density $\left(\rho\right)$,
pressure $\left(P\right)$, and temperature $\left(T\right)$.}

\begin{longtable}{>{\centering}p{3cm}>{\centering}p{3cm}>{\centering}b{3cm}}
\hline \hline $\rho$ & $P$ & $T$\tabularnewline
\endhead
1.50 & 84.00 & 6875\tabularnewline \hline \hline
\endlastfoot
$\left(\textnormal{g/c\ensuremath{m^{3}}}\right)$ &
$\left(\textnormal{GPa}\right)$ &
$\left(\textnormal{K}\right)$\tabularnewline \hline 0.60 & 2.06 &
176\tabularnewline 0.73 & 5.02 & 330\tabularnewline 0.86 & 11.89 &
1441\tabularnewline 1.05 & 25.42 & 2614\tabularnewline 1.16 & 36.74
& 3420\tabularnewline 1.26 & 45.55 & 3620\tabularnewline 1.40 &
60.20 & 4509\tabularnewline
\end{longtable}
\end{table}

To clarify the structural change in methane under shock conditions,
we calculate the PCFs for each pair of atom types, which give the
possibility of finding an atom of a given type at a given distance
from a reference atom. The PCFs, together with atomic structure
along the principal Hugoniot of methane, are presented in Fig. 2. At
the lowest density of
$\rho=0.423\,\textnormal{g/c\ensuremath{m^{3}}}$, the PCF of C-H
bond $g_{\textnormal{C-H}}\left(r\right)$ peaks at about
$1.09\textrm{\AA}$, which corresponds to the equilibrium
internuclear distance of the C-H bond in methane molecule. At this
density, the hydrogen correlation function does not show a maximum
at a distance corresponding to the equilibrium distance of hydrogen
molecule $r_{\textnormal{H-H}}=0.75\textrm{\AA}$. Meanwhile, no
peaks occur in the PCF of C-C bond at $1.50\textrm{\AA}$ or
$1.54\textrm{\AA}$, which corresponds to the typical C-C bond length
in diamond or saturated hydrocarbons. Therefore, the PCFs at
$0.423\,\textnormal{g/c\ensuremath{m^{3}}}$ indicate that methane
remains its ideal molecular configurations without dissociation. As
the density is compressed to
$\rho=1.16\,\textnormal{g/c\ensuremath{m^{3}}}$, methane molecules
dissociate with the increase of density and temperature, which is
indicated in Fig. 2(b) by the significant reduction and broadening
of the maxima of $g_{\textnormal{C-H}}\left(r\right)$ around
$1.09\textrm{\AA}$. On the other hand, the $g_{\textnormal{H-H}}$
and $g_{\textnormal{C-C}}$ PCFs at this density indicate that small
amount of hydrogen and saturated hydrocarbons form molecules upon
dissociation of methane. Consistently, the atomic configuration
consists of a mixture of methane and ethane with the residual
hydrogen atoms both in molecular and atomic forms as shown in the
inset of Fig. 2(b). As the density is further increased to 1.26
$\textnormal{g/c\ensuremath{m^{3}}}$, maximum of the PCF
$g_{\textnormal{C-H}}\left(r\right)$ continues to reduce and broaden
while the peaks of $g_{\textnormal{H-H}}\left(r\right)$ and
$g_{\textnormal{C-C}}\left(r\right)$ increase, which indicates that
methane further dissociates into hydrogen and some hydrocarbons of
higher molecular weight. This trend persists up to a density of 1.4
$\textnormal{g/c\ensuremath{m^{3}}}$, except that a new species of
long alkane chains show up in the atomic configuration. It is
indicative of the formation of C-C bonds favored with increasing
density, though still no diamond-like carbons form in the density
range considered here, in agreement with earlier predictions.

\begin{figure}
\includegraphics[bb=0bp 0bp 539bp 442bp,clip,scale=0.55]{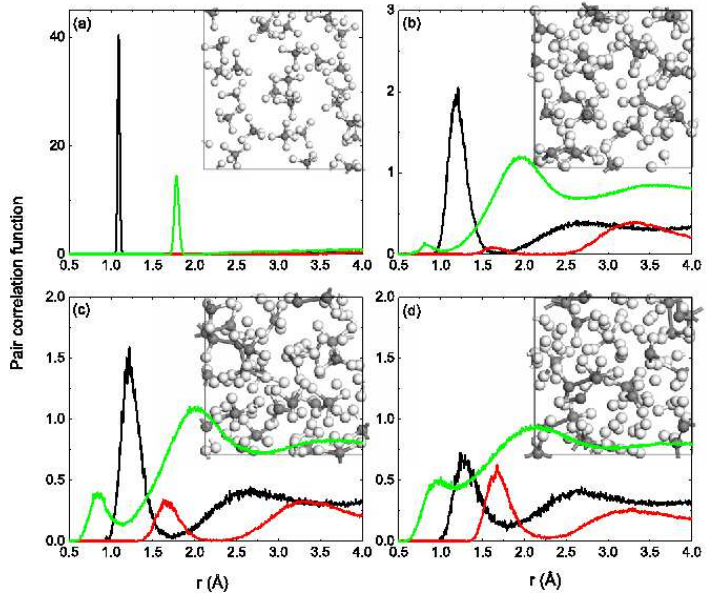}

\caption{Pair-correlation functions for C-C (red or gray line), C-H (black
line), H-H (green or light gray line) along the principal methane
Hugoniot. The atomic structure, where carbon and hydrogen atoms are
denoted by gray and white balls, respectively, is also provided in
the insets. (a)$\rho=0.423\,\textnormal{g/c\ensuremath{m^{3}}}$,
$T=111\,\textnormal{K}$; (b)$\rho=1.16\,\textnormal{g/c\ensuremath{m^{3}}}$,
$T=3420\,\textnormal{K}$; (c)$\rho=1.26\,\textnormal{g/c\ensuremath{m^{3}}}$,
$T=3620\,\textnormal{K}$; (d)$\rho=1.40\,\textnormal{g/c\ensuremath{m^{3}}}$,
$T=4509\,\textnormal{K}$.}

\end{figure}

\section{DYNAMIC OPTICAL and ELECTRONIC PROPERTIES}

Following the QMD simulations, a total number of 10 to 15
configurations are selected from an equilibrated (in an average
sense) portion of the molecular dynamics run, typically sampling the
final picosecond of evolution. The configurations are spaced at time
steps separated by at least the correlation time, the $e-$folding
time of the velocity autocorrelation function. The calculated
properties should be averaged over the number of representative
configurations or snapshots employed. For each of these
configurations, the electrical conductivity is calculated using the
most general formulation given by the Kubo-Greenwood formulation,
without particular assumptions made on the ionic structure or on the
electron-ion interactions. In the framework of the quasi-independent
particle approximation, the Kubo-Greenwood formulation
\cite{Kubo1957,Greenwood1958} gives the real part of the electrical
conductivity as a function of frequency $\omega$,

\begin{eqnarray}
\sigma_{1}\left(\omega\right) & = & \frac{2\pi}{3\omega\Omega}\underset{\mathbf{k}}{\sum}w\left(\mathbf{k}\right)\overset{N}{\underset{j=1}{\sum}}\overset{N}{\underset{i=1}{\sum}}\overset{3}{\underset{\alpha=1}{\sum}}\left[f\left(\epsilon_{i},\mathbf{k}\right)-f\left(\epsilon_{j},\mathbf{k}\right)\right]\nonumber \\
 &  & \times\left|\left\langle \Psi_{j,\mathbf{k}}\right|\nabla_{\alpha}\left|\Psi_{i,\mathbf{k}}\right\rangle \right|^{2}\delta\left(\epsilon_{j,\mathbf{k}}-\epsilon_{i,\mathbf{k}}-\hbar\omega\right),\end{eqnarray}
where $\omega$ is the frequency, $\Omega$ is the volume of the
supercell, $N$ is the total number of energy bands used,
$\Psi_{i,\mathbf{k}}$ and $\epsilon_{i,\mathbf{k}}$ are the
electronic eigenstates and eigenvalues for the electronic state $i$
at $\mathbf{k}$, $f\left(\epsilon_{i},\mathbf{k}\right)$ stands for
the Fermi distribution function, and $w\left(\mathbf{k}\right)$
represents the $\mathbf{k}$-point weighting factor. Other properties
can be directly derived from the frequency-dependent real part of
the electrical conductivity. An application of the Kramer-Kr\"{o}nig
relation yields the imaginary part $\sigma_{2}\left(\omega\right)$
as

\begin{eqnarray}
\sigma_{2}\left(\omega\right) & = &
-\frac{2}{\pi}P\int\frac{\sigma_{1}\left(\nu\right)\omega}{\left(\nu^{2}-\omega^{2}\right)}d\nu,\end{eqnarray}
where $P$ stands for the principal value of the integral. The real
and imaginary parts of dielectric function, in turn, follow
immediately from the electrical conductivity as

\begin{eqnarray}
\epsilon_{1}\left(\omega\right) & = & 1-\frac{4\pi}{\omega}\sigma_{2}\left(\omega\right),\end{eqnarray}

\begin{eqnarray}
\epsilon_{2}\left(\omega\right) & = &
\frac{4\pi}{\omega}\sigma_{1}\left(\omega\right).\end{eqnarray} And
then the real $n\left(\omega\right)$ and imaginary
$k\left(\omega\right)$ parts of the optical refraction index have
relation with the dielectric function by a simple formula,

\begin{eqnarray}
\epsilon\left(\omega\right) & = &
\epsilon_{1}\left(\omega\right)+i\epsilon_{2}\left(\omega\right)=\left[n\left(\omega\right)+ik\left(\omega\right)\right]^{2}.\end{eqnarray}
Finally, the reflectivity $r\left(\omega\right)$ and absorption
coefficient $\alpha\left(\omega\right)$ can be determined from these
quantities,

\begin{eqnarray}
r\left(\omega\right) & = & \frac{\left[1-n\left(\omega\right)\right]^{2}+k\left(\omega\right)^{2}}{\left[1+n\left(\omega\right)\right]^{2}+k\left(\omega\right)^{2}},\end{eqnarray}

\begin{eqnarray}
\alpha\left(\omega\right) & = & \frac{4\pi}{n\left(\omega\right)c}\sigma_{1}\left(\omega\right).\end{eqnarray}

\begin{figure}
\includegraphics[scale=0.75]{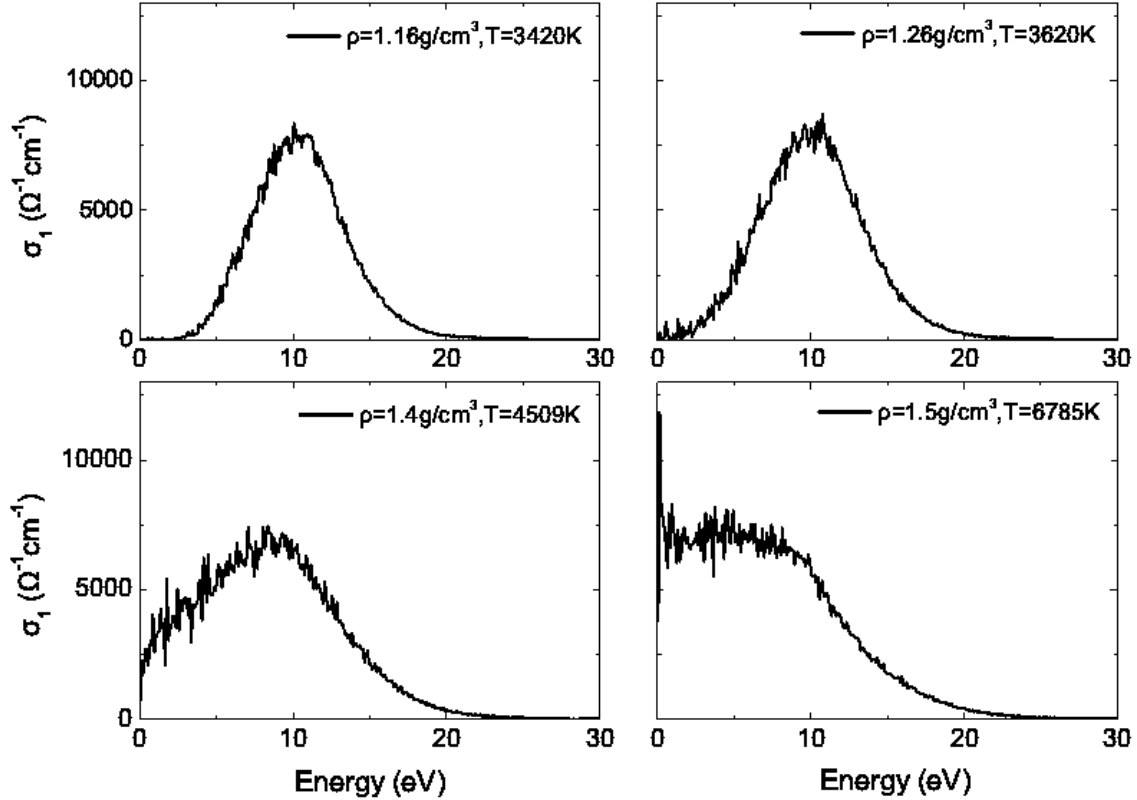}

\caption{The real part of the dynamic electrical conductivity along
the principal Hugoniot. Data have been averaged over 10 to 15
uncorrelated MD configurations.}

\end{figure}

In Fig. 3 the behavior of the frequency-dependent conductivity
$\sigma_{1}\left(\omega\right)$ at different Hugoniot points is
reported in {}``raw'' data form. It is found that
$\sigma_{1}\left(\omega\right)$ at these four different densities
exhibit a uniform feature that they peak around 10.0 eV, which can
be associated with transitions to the lowest excited states, and
vanish above 25.0 eV. With the increase of density and temperature
along the principal Hugoniot, the shape of
$\sigma_{1}\left(\omega\right)$ mostly keeps the same, but the main
peak moves to lower frequencies, which consequently leads to a
significant increase in dc conductivity, given as
$\sigma_{dc}=\underset{\omega\rightarrow0}{lim}\sigma_{1}\left(\omega\right)$.
This significant change of the dc conductivity along principal
Hugoniot is displayed in Fig. 4. For comparison, the experimental
data is also included. The difference between our calculated results
and the experimental results may comes from the under-prediction of
the Hugoniot temperatures for methane by quantum molecular dynamics.
The dc conductivity is negligible for pressures up to 40 GPa, where
methane exhibits insulating behavior. Then, it increases to 128
$\Omega^{-1}\textnormal{c\ensuremath{m^{-1}}}$ for pressures between
40 and 50 GPa. For pressures above 55 GPa, the dc conductivity
increases to a value lager than 1000
$\Omega^{-1}\textnormal{c\ensuremath{\textnormal{m}^{-1}}}$. A
decisive physical assertion to define the metallicity of such
disordered system has been proposed by Ioffe et al. and developed by
Mott \cite{Ioffe1960,Mott1990}, which states that any
high-temperature, disordered system will remain metallic or indeed
attain metallic status if the characteristic mean free path of the
valence electrons exceeds the mean distance between the constituent
atoms or molecules providing those carriers of electrical current.
This simple but powerful argument leads to an estimate of the
minimum metallic conductivity of about 2000
$\Omega^{-1}\textnormal{c\ensuremath{\textnormal{m}^{-1}}}$ for
fluid hydrogen, rubidium, caesium and mercury, whereby transitions
from non-metallic to metallic fluids have recently been observed in
the nominally non-mitallic chemical elements hydrogen, oxygen and
nitrogen at high pressures and temperatures
\cite{Nellis2003,Nellis2004}. With the similar minimum metallic
conductivity for fluid methane, it is indicated that nonmetal-metal
transition takes place in shocked methane. This nonmetal-metal
transition clearly suggests that the dissociation of the molecular
fluid along the Hugoniot has major consequences to the electrical
conductivity of the system. Interestingly, such nonmetal-metal
transition in fluid hydrogen occurs  under similar conditions
\cite{Nellis1992,Holst2008}.

\begin{figure}
\includegraphics[scale=0.75]{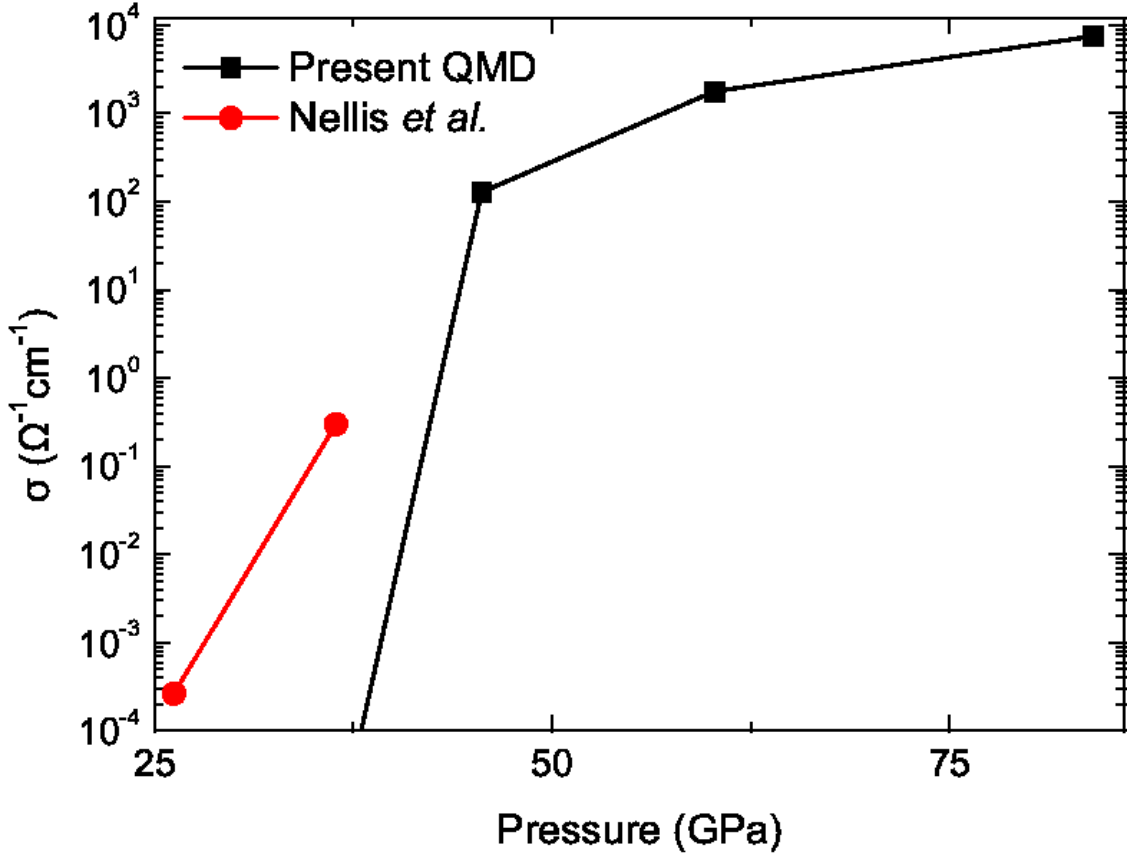}

\caption{The dc conductivity of our results and data from Ref. 8 are
plotted along the principal Hugoniot. }

\end{figure}

As the optical reflectivity of fluid has been successfully probed in
dynamic compression experiments \cite{Collins1998}, we show in Fig.
5 the reflectivity $r\left(\omega\right)$ at typical wavelengths of
404, 808 and 1064 nm spanning the visible spectrum. It can be seen
that a measurable reflectivity arises from 0.05 to 0.35-0.68 for the
principal shocks, which is related with the high-pressure
nonmetal-metal transition. For pressures between 25 and 45 GPa
reflectivity increases monotonically and is essentially equal for
all three frequencies, while around 55 GPa the increase becomes
shaper due to the nonmetal-metal transition, after which the
increase slows down.

\begin{figure}
\includegraphics[scale=0.75]{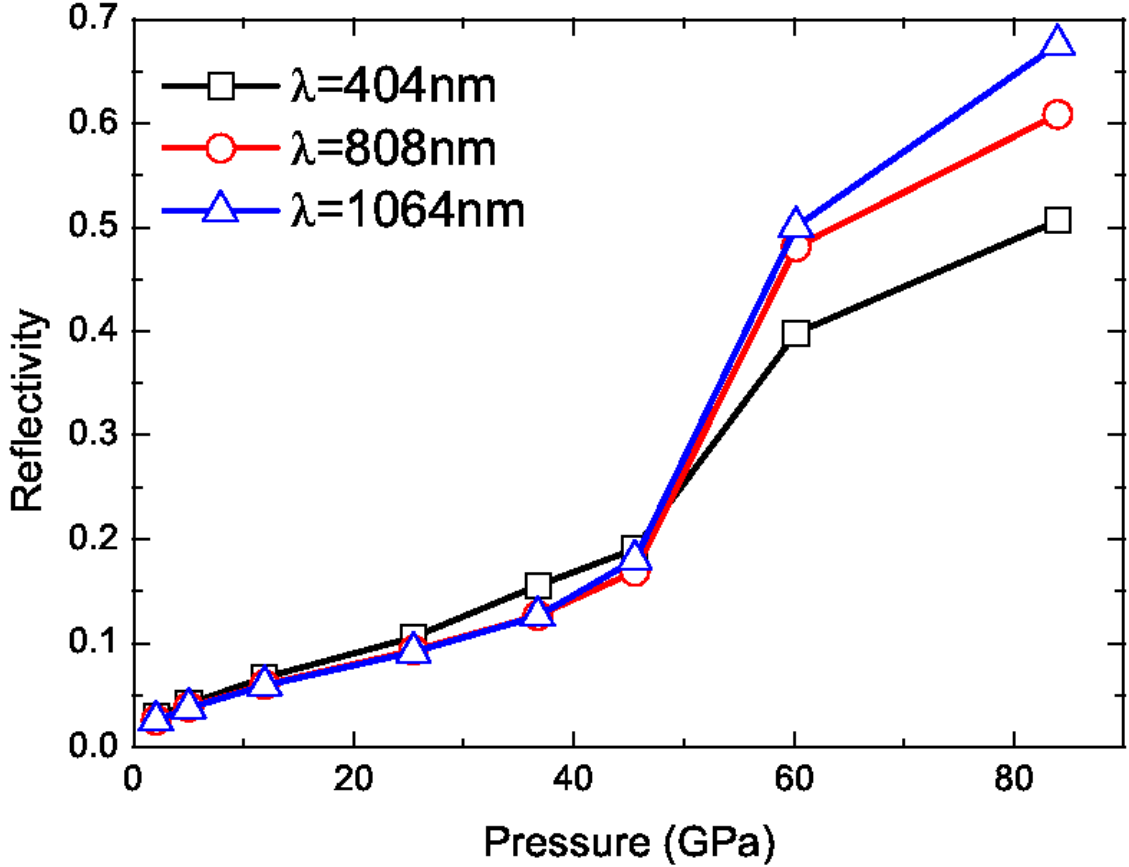}

\caption{Optical reflectivity of shocked methane for wavelengths of 404 (black
square), 808 (red circle) and 1064 nm (blue triangle) along the Hugoniot.}

\end{figure}

Furthermore, we examine the variation of the electron density of
states (EDOS) along the Hugoniot to clarify the origin of this
nonmetal-metal transition. Still use the selected ten to fifteen
configurations used in the above optical properties calculations, we
calculate the electron density of states. Each of these
configurations repeats periodically throughout the space and forms
the elements of the calculation. In Fig. 6 the averaged EDOS is
presented for different Hugoniot points. As can be seen, at
$\rho=1.16\ \textnormal{g/c\ensuremath{m^{3}}}$ and
$T=3420\,\textnormal{K}$, EDOS clearly exhibits a gap around 2.4 eV,
where only thermally activated electron transport occurs as in
semiconductors. With the density and temperature increasing, the
band gap is reduced gradually and closed eventually. And thus a
higher, metal-like conductivity follows.

\begin{figure}
\includegraphics[scale=0.35]{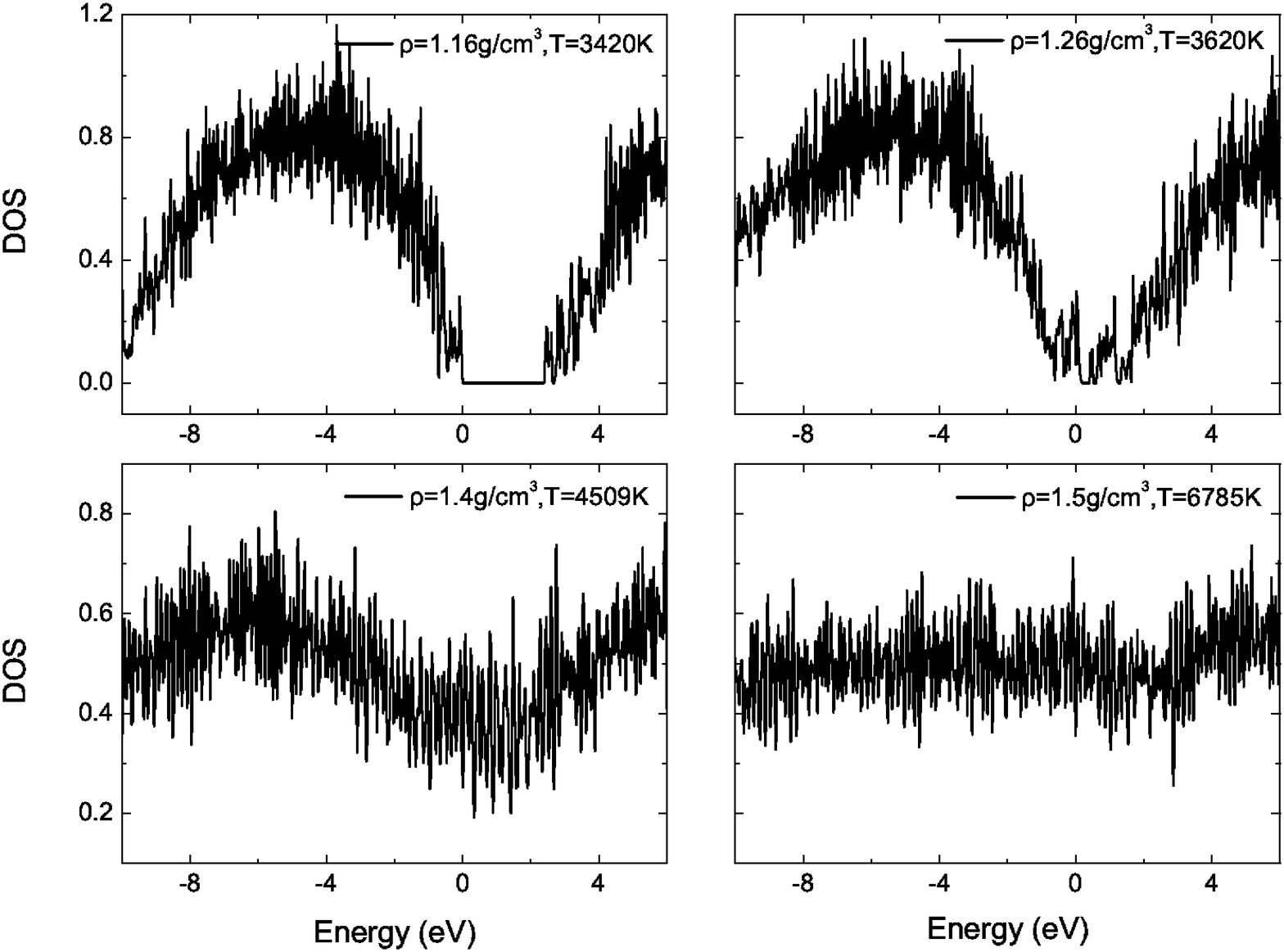}

\caption{Electronic density of states of liquid methane along the
principal Hugoniot:
(a)$\rho=1.05\,\textnormal{g/c\ensuremath{m^{3}}}$,
$T=2614\,\textnormal{K}$;
(b)$\rho=1.16\,\textnormal{g/c\ensuremath{m^{3}}}$,
$T=3420\,\textnormal{K}$;
(c)$\rho=1.26\,\textnormal{g/c\ensuremath{m^{3}}},\:
T=3620\,\textnormal{K}$;
(d)$\rho=1.40\,\textnormal{g/c\ensuremath{m^{3}}},\:
T=4509\,\textnormal{K}$. Data have been averaged over 10 to 15
uncorrelated MD configurations. The zero of the energy scale shows
the position of the Fermi level.}

\end{figure}

Another way to characterize this behavior is the change of the
charge density with increasing density as shown in Fig. 7, from
which it is clearly seen that a transition from a disconnected
network (left panel) to a connected one (right) occurs as density
increases. This suggests that the electrons are still localized for
the lower density, while transient filaments and lusters form as
percolating path for the higher density, which leads to a metal-like
behavior.

\begin{figure}
\includegraphics[bb=0bp 0bp 568bp 234bp,clip,scale=0.5]{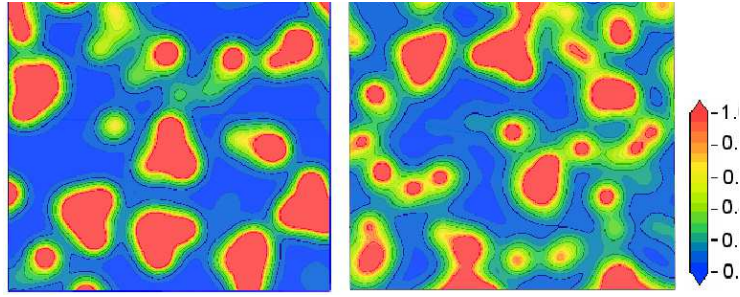}

\caption{Typical contour plots of the charge densities in units of
$\textnormal{e}/\textrm{\AA}^{3}$ for 0.86
$\textnormal{g/c\ensuremath{m^{3}}}$ at 1441 K (left panel) and 1.4
$\textnormal{g/c\ensuremath{m^{3}}}$ at 4509 K (right panel) along a
slice through the simulation box.}

\end{figure}

\section{CONCLUSIONS}

In summary, through systematic QMD simulations we have determined
the shocked EOS and clarified the high-pressure nonmetal-metal
transition of the fluid methane. The increase of electrical
conductivity with pressure can be ascribed to the dissociation of
the molecular fluid. However, no diamond-like carbon forms in the
density range considered here. In addition, the optical properties
of warm dense methane are also calculated, from which an
experimentally measurable increase in the reflectivity associated
with the high-pressure nonmetal-metal transition has been found. It
is expected that our simulated results of shocked methane would have
a strong influence on models for planetary interiors.
\begin{acknowledgments}
This work was supported by NSFC under Grant No. 51071032 and
10734140, and by the National Fundamental Security Research Program
of China.\end{acknowledgments}

\end{document}